\documentclass[aps,apl,preprint,showpacs,groupedaddress,doublespace]{revtex4-1}  % for double-spaced preprint
\usepackage{graphicx}  % needed for figures
\usepackage{amsmath}   % for math
\usepackage{amssymb}   % for math
\usepackage{bm}        % for math

\begin{document}
\newcommand{\etal}{{\em et al.}}{}
\newcommand{\fig}[1]{Fig.~\ref{#1}}
\newcommand{\tab}[1]{{Table ~\ref{#1}}}
\newcommand{\AN}{$\rm{\AA}$\,}{}
\newcommand{\ww}{.5}

\title{Mechanics properties of Mono-layer Hexagonal Boron Nitride: Ab initio study}
\author{Qing Peng, Amir Zamiri and Suvranu De}
\affiliation{
\begin{tabular}{cl}
 & Department of Mechanical, Aerospace and Nuclear Engineering,\\& Rensselaer Polytechnic Institute, Troy, NY 12180, U.S.A.\\
\end{tabular}
}

\begin{abstract}
We introduced a method to obtain the continuum description of the elastic properties of monolayer h-BN
through {\em ab initio} density functional theory.
This thermodynamically rigorous continuum description                                
of the elastic response is formulated by expanding the
elastic strain energy density in a Taylor series in strain
truncated after the fifth-order term. 
we obtained a total of fourteen nonzero independent
elastic constants for the up to tenth-order tensor.
We predicted the pressure dependent second-order elastic moduli.
This continuum formulation is suitable for
incorporation into the finite element method.
\end{abstract}

 \maketitle

\section{Introduction}
The area of research on 2D nanomaterials with potential next generation device application 
has seen tremendous progress in the past few recent years. 
An example of such nanostructures is hexagonal boron nitride (h-BN) monolayer 
which is analog of graphene having a honeycomb lattice structure \cite{ISI:000275858200036}. 
Hexagonal boron nitride is chemically inert, has a high thermal conductivity, 
and is highly temperature resistant to oxidation. 
Due to its outstanding properties, h-BN has found wide applications 
in micro and nano-devices such as insulator with high thermal 
conductivity in electronic devices \cite{ISI:000275858200036}, 
ultraviolet-light emitter in optoelectronics \cite{ISI:000073632900036,ISI:000220586800019,ISI:000172400600026,ISI:A1994PT46600007,ISI:000221890700023}, 
and as nano-fillers in high strength and thermal conductive 
nanocomposites \cite{ISI:000236654400005,ISI:000268652700007}. 
In very recent works, it has been shown that a tunable band gap nanosheet 
can be constructed by fabrication of hybrid nanostructures made of 
graphene/h-BN domains which opens a new venue for huge research 
in the application of h-BN for electronics\cite{PhysRevB.76.073103,ISI:000276953500024}. 

Given the aforementioned potential applications, 
the complete knowledge of mechanical and physical properties of h-BN 
monolayer, however, is still lacking. 
The previous primary works have reported that h-BN monolayer 
has a bulk modulus around 160 Pa.m and a bending modulus 
around 31.2GPa (Ref. \cite{ISI:000275858200036,ISI:000269356900019,ISI:000278888600004,PhysRevB.79.115442}) 
(-- Amir, could you please double check this? It does not make sense.) 
Several experimental and atomistic simulation studies, 
mostly on graphene and carbon nanotubes,  have probed that 2D nanosheets and nanotubes usually 
show a nonlinear elastic deformation during the tension up to the intrinsic 
strength of the material followed by a strain softening 
up to the fracture \cite{ISI:000257713900044,PhysRevB.76.064120,ISI:000221843400091,PhysRevB.75.075412,ISI:000207921500004}. 
To establish a continuum based framework to capture this nonlinear 
elastic behavior of the 2D nanosheets, the higher order of the 
elastic constants must be considered in the 
strain energy density function \cite{PhysRevLett.102.235502,PRBwei2009}. 
In such a model, the strain energy density is expanded in 
a Taylor series to include both quadratic as well as higher order terms in strain. 
The quadratic term accounts for the linear elastic response of 
the material while the cubic and higher order terms account the strain 
softening of the elastic stiffness. 
The higher order terms also can be used to define other anharmonic properties 
of this 2D nanostructure including phenomena such as thermal expansion, 
phonon-phonon interaction, etc \cite{ISI:A1981LZ61200003}.

The goal of this paper is to find the continuum description of the elastic properties of monolayer h-BN.
To achieve that, we first examine the elastic properties of h-BN monolayers using {\em ab initio} density functional theory. 
We adopt a fifth-order series expansion of
the strain energy density function in order to model the inplane 
elastic properties of h-BN and demonstrate that the
resulting continuum description ͑now with fourteen independent 
elastic constants͒ describes accompanying ab initio DFT calculations 
with high accuracy in the infinitesimal strain regime as
well as at finite strains, including the strain at the intrinsic
stress and beyond. A higher rank tensor is associated with
each term of the series expansion and the components of the
tensor represent the continuum elastic properties. 
Previous authors had determined the nonzero independent tensor
components that correspond to the symmetry elements of
graphene for the second-, third-, fourth-order terms and fifth-order term from stress-strain response for graphene \cite{PRBwei2009}. 
We extended the method with least-squares solution to over-determined (up to eighth-rank tensor) 
and well-determined (tenth-rank tensor) linear equations. 
We applied this advanced method to obtain the continuum description of the elastic properties of of monolayer h-BN in the following sections.

\section{Nonlinear elasticity theory applied to 2D hexagonal structure}

\begin{figure}
\includegraphics[width=\ww\textwidth]{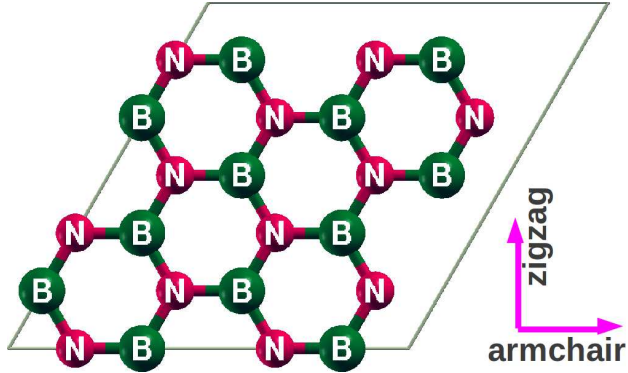}
\caption
{\label{fig:config} Atomic super cell (24 atoms) of h-BN in the undeformed reference configuration.} 
\end{figure}

We used a super cell containing 12 B and 12 N atoms in one plane, with periodic boundary conditions.  
The undeformed reference configuration  is shown in \fig{fig:config}, with lattice vectors $\mathbf{H_i}$, $i=1,2,3$.
When a macroscopically homogeneous deformation (deformation gradient tensor \cite{NLFEM} $\mathbf{F}$) applied, 
the lattice vectors of the deformed h-BN are $\mathbf{h}_i=\mathbf{F} \mathbf{H}_i$. 
The Lagrangian strain \cite{LagrangianStrain} is defined as {\boldmath$\eta$}=$\dfrac{1}{2}(\mathbf{F^T}\mathbf{F}-\mathbf{I})$, 
where $\mathbf{I}$ is the identity tensor.
The strain energy density has functional form of $\mathbf{\Phi}=\mathbf{\Phi}$({\boldmath{$\eta$}}). 
The elastic properties of a
material are determined from $\mathbf{\Phi}$, which is quadratic in strain for a linear elastic material. 
Nonlinear elastic constitutive behavior is established
by expanding $\mathbf{\Phi}$ in a Taylor series in terms of powers of
strain {\boldmath$\mathbf{\eta}$}. 
The symmetric second Piola-Kirchhoff
stress tensor, $\Sigma_{ij}$, can be expressed (up to fifth order) as \cite{PRBwei2009}: 
\begin{align}
\label{eq:sigma}
\Sigma_{ij} &=\frac{\partial \Phi}{\partial \eta_{ij}} \nonumber \\
            &=C_{ijkl}\eta_{kl} +\frac{1}{2!}C_{ijklmn}\eta_{kl}\eta_{mn} \nonumber \\
            &+\frac{1}{3!}C_{ijklmnop}\eta_{kl}\eta_{mn}\eta_{op} \nonumber \\
            &+\frac{1}{4!}C_{ijklmnopqr}\eta_{kl}\eta_{mn}\eta_{op}\eta_{qr}.
\end{align}
where $\eta_{ij}$ is Lagrangian elastic strain. Summation convention
is employed for repeating indices; lower case subscripts
range from 1 to 3. Herein $\mathbf{C}$ denotes each higher-order elastic
modulus tensor; the rank of each tensor corresponds to the
number of subscripts.
The second-order elastic constants
(SOEC͒), $C_{ijkl}$, third-order elastic constants (TOEC͒), $C_{ijklmn}$,
fourth-order elastic constants (FOEC͒), $C_{ijklmnop}$, and fifth-
order elastic constants (FFOEC͒), $C_{ijklmnopqr}$, are given by the
components of the fourth-, sixth-, eighth-, and tenth-rank
tensors, respectively. 

We used conventional Voigt notation \cite{Voigt} for subscripts: 11 $\rightarrow$ 1, 22
$\rightarrow$ 2, 
33 $\rightarrow$ 3, 23 $\rightarrow$ 4, 31 $\rightarrow$ 5, and 12 $\rightarrow$ 6. 
Please note that for strain $\eta_{4}=2\eta_{23}$, $\eta_{5}=2\eta_{31}$, $\eta_{6}=2\eta_{12}$.
Eqs. (1) can be rewritten as 
\begin{align}
\label{eq:sigmaI}
\Sigma_{I}&=\frac{\partial \Phi}{\partial \eta_{I}}=C_{IJ}\eta_{J} +\frac{1}{2!}C_{IJK}\eta_{J}\eta_{K} \nonumber \\
&+\frac{1}{3!}C_{IJKL}\eta_{J}\eta_{K}\eta_{L}+\frac{1}{4!}C_{IJKLM}\eta_{J}\eta_{K}\eta_{L}\eta_{M}.
\end{align}
where the summation convention for upper case subscripts
runs from 1 to 6.

In this study, we modeled the monolayer h-BN as two
dimensional (2D͒) structure and assume that the deformed state of the
monolayer h-BN is such that the contribution of bending to the strain
energy density is negligible as compared to the in-plane
strain contribution. 
This assumption is reasonable since the radius of curvature
of out-of-plane deformation are significantly larger than
the in-plane inter-atomic distance. 
Then the stress state of
monolayer h-BN under those assumptions can be assumed to be 2D and  we only consider the in-plane
stress and strain components for these kind of structures.

The components of the TOEC, FOEC, and
FFOEC tensors can be determined based on the symmetries
of the graphene atomic lattice ͑point group $D_{6h}$ which con-
sists of a sixfold rotational axis and six mirror planes as formulated in ref \cite{PRBwei2009}.

The fourteen independent elastic constants of h-BN
are determined by a least-squares fit to stress-strain results
 from ab initio DFT simulations in two steps. At the first step, 
we use least-squares fit to five stress-strain responses. 
Five relationships between stress
and strain are necessary because there are five independent
FFOECs. We obtain the stress-strain relationships by simu-
lating the following deformation states: uniaxial strain in the
zigzag direction; uniaxial strain in the armchair
direction; and, equibiaxial strain. 
From the first step, we the components of SOEC, TOEC, FOEC are over-determined 
(i.e, the number of linearly independent variables are greater than the number of constrains),
and the FFOEC are well-determined (the number of linearly independent variables are equal to the number of constrains).
Under such circumstance, the second step is needed: least-square solution to these over- and well- determined linear equations.

At the first step, we carried out three deformations, uniaxial strain in the zigzag direction (case $z$), 
uniaxial strain in the armchair direction (case $a$) and equibiaxial strain (case $b$). 
For uniaxial strain in the zigzag direction, the strain tensor is,
\begin{equation}
\label{eq:eta}
\eta_{ij}^z=
\begin{bmatrix} 0& 0 & 0 \\0 & \eta_z & 0 \\0 & 0 & 0 \end{bmatrix},
\end{equation}
where $\eta_z$ is the amount of strain in zigzag direction.

For a given strain tensor, the associated deformation gradient tensor is not unique.
The various possible solutions differing from one to another by a rigid rotation.
Here the lack of a one-to-one map relationship between the strain tensor
and deformation gradient tensor is not concern since the calculated energy is invariant under
rigid deformation \cite{Zhou20081609,WangRui2010}. One of the corresponding deformation gradient
tensor $\mathbf{F}_{z}$ for uniaxial strain in the zigzag direction 
is selected as
\begin{equation}
\mathbf{F}_{z}=
\begin{bmatrix} 1 & 0 & 0 \\0 & \varepsilon_z & 0 \\0 & 0 & 1 \end{bmatrix},
\end{equation}
where $\varepsilon_z$ is the stretch ration $\varepsilon$ in the zigzag direction. 
$\varepsilon$ is determined by the Lagrangian elastic strain through equation
\begin{equation}
\frac{1}{2}\varepsilon^2+\varepsilon-\eta=0.
\end{equation}

The stress-strain relationships of the uniaxial strain in the zigzag direction are 
\begin{equation}
\label{eq:sz1}
\Sigma^z_1=C_{12}\eta_z+\dfrac{1}{2}C_{112}\eta^2_z+\dfrac{1}{6}C_{1112}\eta^3_z+\dfrac{1}{24}C_{11112}\eta^4_z,
\end{equation}
\begin{equation}
\label{eq:sz2}
\Sigma^z_2=C_{11}\eta_z+\dfrac{1}{2}C_{111}\eta^2_z+\dfrac{1}{6}C_{1111}\eta^3_z+\dfrac{1}{24}C_{11111}\eta^4_z,
\end{equation}

For uniaxial strain in the armchair direction, the strain tensor is,
\begin{equation}
\eta_{ij}^a=
\begin{bmatrix} \eta_a &0 & 0 \\0 & 0 & 0 \\0 & 0 & 0 \end{bmatrix},
\end{equation}

One of the corresponding deformation gradient
tensor $\mathbf{F}_{a}$ for uniaxial strain in the armchair direction is 
\begin{equation}
\mathbf{F}_{a}=
\begin{bmatrix} \varepsilon_a &0 & 0 \\0 & 1 & 0 \\0 & 0 & 1 \end{bmatrix},
\end{equation}
where $\varepsilon_a$ is $\varepsilon$ in the armchair direction.
The stress-strain relationships are 
\begin{align}
\label{eq:sa1}
\Sigma^a_1 &=C_{11}\eta_a+\dfrac{1}{2}C_{222}\eta^2_a+\dfrac{1}{6}C_{2222}\eta^3_a+\dfrac{1}{24}C_{22222}\eta^4_a,
\end{align}
\begin{align}
\label{eq:sa2}
\Sigma^a_2 &=C_{12}\eta_a+\dfrac{1}{2}(C_{111}-C_{222}+C_{112})\eta^2_a \nonumber \\
           &+\dfrac{1}{12}(C_{1111}+2C_{1112}-C_{2222})\eta^3_a+\dfrac{1}{24}C_{12222}\eta^4_a,
\end{align}

For equibiaxial strain in-plane, $\eta_a=\eta_z=\eta$,the strain tensor is,
\begin{equation}
\eta_{ij}^b=
\begin{bmatrix} \eta & 0 & 0 \\0 & \eta & 0 \\0 & 0 & 0 \end{bmatrix},
\end{equation}
The corresponding deformation gradient
tensor $\mathbf{F}_{b}$ for equibiaxial strain in-plane is 
\begin{equation}
\mathbf{F}_{b}=
\begin{bmatrix} \varepsilon & 0 & 0 \\0 & \varepsilon & 0 \\0 & 0 & 1 \end{bmatrix}.
\end{equation}.
The stress-strain relationships are
\begin{align}
\label{eq:sb}
\Sigma^b_1 &=\Sigma^b_2=(C_{11}+C_{12})\eta+\frac{1}{2}(2C_{111}-C_{222}+3C_{112})\eta^2 \nonumber \\
           &+\frac{1}{6}(\frac{3}{2}C_{1111}+4C_{1112}+3C_{1122}-\frac{1}{2}C_{2222})\eta^3 \nonumber \\
           &+\frac{1}{24}(3C_{11111}+10C_{11112}-5C_{12222}+10C_{1122} \nonumber \\
           &-2C_{22222})\eta^4.
\end{align}
All fourteen elastic constants contribute to the expressions
for stress-strain response for these three deformation states.
But the components of SOEC, TOEC, FOEC are over-determined.
As the second step, we apply the least-square solution to these over- and well- determined linear equations.
We used least-squares solutions to solve the equations
$\mathbf{A} \cdot \mathbf{C}=\mathbf{\Sigma}$ by computing the elastic constants that minimizes the Euclidean 2-norm 
$\|\mathbf{\Sigma}-\mathbf{A}\cdot\mathbf{C}\|^2$. For SOEC components, $C_{11},C_{12}$ is obtained by
\begin{equation}
\label{eq:c2}
\begin{bmatrix} 1 & 0 \\0&1 \\0&1 \\1 &0 \\ 1& 1 \end{bmatrix} 
\begin{bmatrix} C_{11}\\C_{12} \end{bmatrix} 
=
\begin{bmatrix} \Sigma^z_2(O_1) \\ \Sigma^z_1(O_1) \\\Sigma^a_2(O_1) \\\Sigma^a_1(O_1) \\\Sigma^b_1(O_1) \end{bmatrix} 
\end{equation}
where $\Sigma^z_1(O_1)$ is the coefficient of the first order of strain in $\Sigma^z_1$ (Eqn.9). Similar notation for the others. 
The Young's modulus is $E=(C_{11}^2-C_{12}^2)/C_{11}$ and Poisson's ration is $\nu=C_{12}/C_{11}$.

For TOEC components $C_{111}, C_{112} and C_{222}$ are obtained by
\begin{equation}
\label{eq:c3}
\frac{1}{2}
\begin{bmatrix} 1 & 0 &0\\0&1&0 \\1&1&-1 \\0&0&1 \\ 2&3&-1 \end{bmatrix} 
\begin{bmatrix} C_{111}\\C_{112}\\C_{222} \end{bmatrix} 
=
\begin{bmatrix} \Sigma^z_2(O_2) \\ \Sigma^z_1(O_2) \\\Sigma^a_2(O_2) \\\Sigma^a_1(O_2) \\\Sigma^b_1(O_2) \end{bmatrix} 
\end{equation}
For FOEC components $C_{1111}$, $C_{1112}$, $C_{1122}$ and $C_{2222}$ are  obtained by
\begin{equation}
\label{eq:c4}
\frac{1}{6}
\begin{bmatrix} 1&0&0&0\\0&1&0&0 \\0.5&1&0&-0.5 \\0&0&0&1 \\ 1.5&4&3&-0.5 \end{bmatrix} 
\begin{bmatrix} C_{1111}\\C_{1112}\\C_{1122}\\C_{2222} \end{bmatrix} 
=
\begin{bmatrix} \Sigma^z_2(O_3) \\ \Sigma^z_1(O_3) \\\Sigma^a_2(O_3) \\\Sigma^a_1(O_3) \\\Sigma^b_1(O_3) \end{bmatrix} 
\end{equation}

For FFOEC components $C_{11111}$, $C_{11112}$, $C_{11122}$, $C_{12222}$ and $C_{22222}$ are obtained by
\begin{equation}
\label{eq:c5}
\frac{1}{24}
\begin{bmatrix} 1&1&0&0&0\\0&0&0&0&0 \\0&0&0&1&0 \\0&0&0&0&1 \\ 3&10&10&-5&-2 \end{bmatrix} 
\begin{bmatrix} C_{11111}\\C_{11112}\\C_{11122}\\C_{12222}\\C_{22222} \end{bmatrix} 
=
\begin{bmatrix} \Sigma^z_2(O_4) \\ \Sigma^z_1(O_4) \\\Sigma^a_2(O_4) \\\Sigma^a_1(O_4) \\\Sigma^b_1(O_4) \end{bmatrix} 
\end{equation}

\section{DENSITY-FUNCTIONAL CALCULATIONS}
The stress-strain relationship of graphene under the desired deformation configurations 
is characterized via {\em ab initio} calculations with the density-functional theory (DFT).
DFT calculations were carried out with the Vienna Ab-initio
Simulation Package (VASP)
\cite{VASPa,VASPb,VASPc,VASPd} which is based
on the Kohn-Sham Density Functional Theory (KS-DFT) \cite{DFTa,DFTb} with the generalized gradient approximations as parameterized by Perdew, Burke and Ernzerhof (PBE)  for exchange-correlation functions \cite{GGA}.
The electrons explicitly included in the calculations are  
the ($2s^22p^1$) electrons of boron and ($2s^22p^3$) electrons of nitrogen. 
The core electrons ($1s^2$) of boron and nitrogen are replaced by the projector augmented wave (PAW) and pseudo-potential approach\cite{paw1,paw2}. 
A plane-wave cutoff of 520 eV is used in all the calculations. The calculations are performed at zero temperature.

The criterion to stop the relaxation of the electronic degrees of freedom is set by total energy change to be smaller than 0.000001 eV. 
The optimized atomic geometry was achieved through minimizing Hellmann-Feynman forces 
acting on each atom  until the maximum forces on the ions were smaller than 0.001 eV/\AA.

The atomic structures of all the deformed and undeformed configurations are obtained by fully relaxing a 24-atom-unit cell where all atoms were placed in one plane. The simulation invokes periodic boundary conditions for the two
in-plane directions while the displacement to out-of-plane direction is forbidden. 

The irreducible Brillouin Zone was sampled with a Gamma-centered $19 \times 19 \times 1$ $k$-mesh. Such large $k$-mesh was used to reduce the numerical errors caused by the strain of the systems. The initial charge densities were taken as a superposition of atomic charge densities. 
There was a 14 \AA \, thick vacuum region to reduce the inter-layer interaction to model the single layer system. The results of the calculations are independent of the precise value of the out-of-plane thickness, so there is no
physical interpretation attached to the quantity.

The VASP simulation calculates the true or Cauchy stress,
{\boldmath $\sigma$}, which for monolayer h-BN must be expressed as a 2D force per
length with units of N/m by taking the product of the Cauchy
stress (with units of N/m$^2͒$) and the super-cell thickness of
14 \AA. The Cauchy stress is related to the second Piola-
Kirchhoff(PK͒2) stress $\mathbf{\Sigma}$ as
{\boldmath
\begin{equation}
\mathbf{\Sigma}=J\mathbf{F}^{-1} \sigma(\mathbf{F}^{-1})^T
\end{equation} }
where $J$ is the determinant of the deformation gradient tensor $\mathbf{F}$.

\section{Results and Analysis}

We first optimize the equilibrium lattice constant for monolayer h-BN. 
The total energy as a function of lattice spacing is
obtained by specifying several lattice constants varying
around 1.45 \AA with full relaxations of all the atoms. A least-squares fit of the energy vs lattice
constant with a fourth-order polynomial function yields the
equilibrium lattice constant, a0 = 1.4503 Å, which corre-
sponds to the minimum total energy. 
The result is in good agreement 
with experiments\cite{PhysRevB.68.104102} in h-BN (2.51 \AA)
This most energy favorite structure is set as the strain-free structure in this study and the geometry is shown in \fig{fig:config}.

When the strains are applied, all the atoms are allowed full freedom of motion within plane. A quasi-Newton algorithm is
used to relax all atoms into equilibrium positions within the
deformed unit cell that yields the minimum total energy for
the imposed strain state of the super cell. 

\begin{figure}
\includegraphics[width=\ww\textwidth]{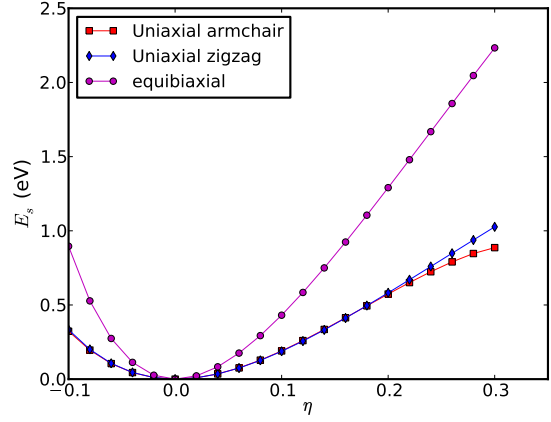}
\caption{\label{fig:eee} 
Energy-strain responses for uniaxial strain in armchair and zigzag directions, 
and equibiaxial strains. 
}              
\end{figure}

Both compression and tension are considered here in order to sampling larger elastic deformation region. 
We studied the behavior of the system under the Lagrangian strain ranged from -0.1 to 0.3 with a increment of 0.02 in each step 
for all three cases. There are 63 {\em ab initio} DFT calculations in total.

The system's energy will increase when strains are applied. 
Here we define strain energy per atom $E_s=(E_{tot}-E_0)/n$, 
where $E_{tot}$ is the total energy of the strained system, 
$E_{0}$ is he total energy of the strain-free system, 
$n$ is the number of atoms in the system. 
\fig{fig:eee} shows the $E_s$ as a function of strain in uniaxial armchair, uniaxial zigzag and equibiaxial deformation. 
$E_s$ responses differently at different strain direction, consistent to the non-isotropic structure of the monolayer h-BN.
$E_s$ are non-symmetrical for compression ($\eta<0$) and tension ($\eta>0$) for all three cases. 
This non-symmetry indicates the anhomonicity of the monolayer h-BN structures. 
$E_s$ deviated from quadratic relationship with $\eta$ at strain of 0.04 in the three tested deformations. 

\begin{figure}
\includegraphics[width=\ww\textwidth]{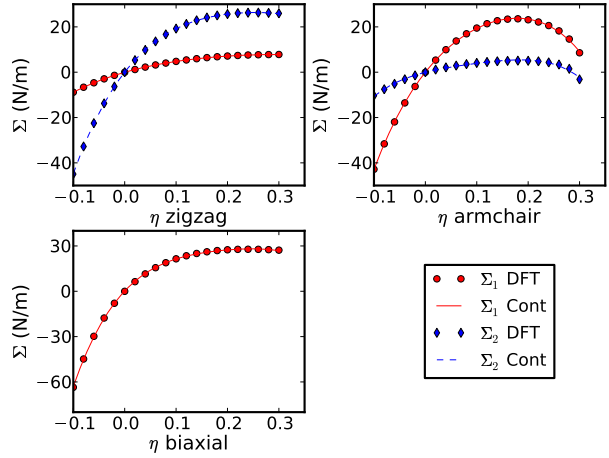}
\caption{\label{fig:elast} Stress-strain responses for uniaxial strain in armchair and zigzag directions, and equibiaxial strains. The continuum responses are the least-square fit of the {\em ab initial} DFT calculations.} 
\end{figure}

The stress (second P-K stress) strain (Lagrangian strain) relationship 
for uniaxial strain in armchair and zigzag directions, 
and equibiaxial strains are shown in \fig{fig:elast}. 
These stress-strain curves reflects the facts of the non-isotropic h-BN structure and anharmonic response in compression and tension. 
For the uniaxial deformation along armchair direction, the maximum stress of $\Sigma_1^a=5.08$, $\Sigma_1^a=23.56$ (N/m) at $\eta_a=0.18$. 
For the uniaxial deformation along armchair direction, 
the maximum stress of $\Sigma_1^z=26.26$ (N/m) at $\eta_z=0.26$ 
and $\Sigma_2^z=7.82$ (N/m) at $\eta_z=0.30$.
For the equibiaxial  deformation, 
the maximum stress of $\Sigma_1^b=\Sigma_2^b=27.81$ (N/m) at $\eta=0.24$.

The elastic constants are the continuum description of the elastic properties. Once we know the elastic constants, we can easily apply it for the continuum description. 
The continuum responses are the least-square fit to 
the stress-stain results from the {\em ab initial} DFT calculations, 
as plotted in \fig{fig:elast}, 
by the equations of (\eqref{eq:sz1},\eqref{eq:sz2},\eqref{eq:sa1},\eqref{eq:sa2},\eqref{eq:sb}).  
We then have 20 values for the fourteen independent elastic constants of h-BN from {\em ab initio} DFT calculations.  The fourteen independent elastic constants of h-BN
are finally determined by solving equations from Eqn. (\eqref{eq:c2},\eqref{eq:c3},\eqref{eq:c4},\eqref{eq:c5}).  
The results of these fourteen independent elastic constants are 
grouped in SOEC, TOEC, FOEC and FFOEC and listed in \tab{tab:a0}.
The in-plane Young's modulus $Y_s=279.2$ (N/m)  and Poison's ratio $\nu=0.2176$  
were obtained from $C_{11}$ and $C_{12}$. 
Our results of $C_{11}$, $C_{12}$, $Y_s$ and $\nu$ are comparable with {\em ab initio} predictioin \cite{PhysRevB.64.235406} and tight-binding calculations\cite{ISI:000073632900036} of BN nanotubes.

\begin{table}
\caption{\label{tab:a0} Nonzero independent components for the SOEC, TOEC, FOEC and FFOEC tensor components, Poisson's ration $\nu$ and in-plane stiffness $Y_s$ of h-BN from DFT calculations.}
\begin{tabular}{|c|c|c|c|}
\hline
SOEC&TOEC&FOEC&FFOEC \\
(N/m)&(N/m)&(N/m)&(N/m)\\
\hline
$C_{11}$=293.1&$C_{111}$=-2515.8&$C_{1111}$=18161&$C_{11111}$=-65265 \\
$C_{12}$=63.76&$C_{112}$=-428.5 &$C_{1112}$=5836 &$C_{11112}$=-8454 \\
              &$C_{222}$=-2300.6&$C_{1122}$=-2868&$C_{11122}$=-67857 \\
$Y_s$=279.2   &                 &$C_{2222}$=12451&$C_{12222}$=-10780 \\
 &                 &                &$C_{22222}$=-117409 \\
\hline
\end{tabular}
\end{table}

\begin{figure}[htp]
\includegraphics[width=\ww\textwidth]{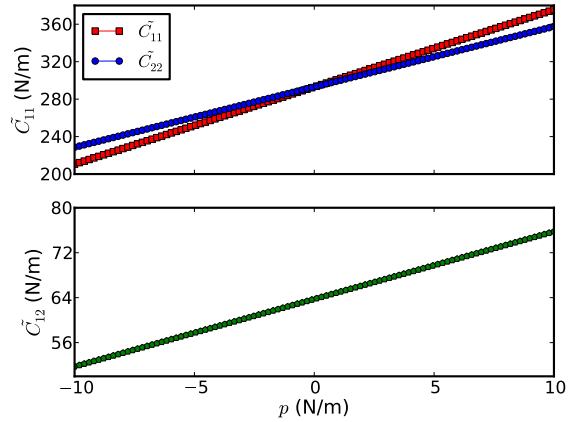}
\caption{\label{fig:pp} Predicted second-order elastic moduli varies with the pressure $p$ acting in the plane of monolayer of h-BN sheet}
\end{figure}

The knowledge of these high order elastic constants is very useful to understand the anharmonicity. 
With the high order elastic constants, we can easily study the second-order elastic moduli 
on the pressure $p$ acting in the plane of monolayer of h-BN sheet. 
Explicitly, while the pressure is applied, the second-order elastic moduli are transformed 
according to the relationships \cite{Voigt,ISI:000288175100035}:
\begin{align}
\label{eq:c2a}
\tilde{C_{11}}=C_{11}-(C_{111}+C_{112})\frac{1-\nu}{Y_s}p,
%\tilde{C_{22}}=C_{11}-C_{222}\frac{1-\nu}{Y_s}p \\
%\tilde{C_{12}}=C_{12}-C_{112}\frac{1-\nu}{Y_s}p
\end{align}
\begin{equation}
\label{eq:c2a}
\tilde{C_{22}}=C_{11}-C_{222}\frac{1-\nu}{Y_s}p,
\end{equation}
\begin{equation}
\label{eq:c2a}
\tilde{C_{12}}=C_{12}-C_{112}\frac{1-\nu}{Y_s}p.
\end{equation}
The second-order elastic moduli increase linearly with the applied pressure $p$ within the third-order term trucation, as demonstrated in \fig{fig:pp}. 
These equations and plots also indicate that the h-BN layer respond to compression or tension along different directions in different manners. 
While pressure presented, the $\tilde{C_{11}}$ is not symetrical to $\tilde{C_{22}}$ any more, although the difference is relatively small.
Olny when $p=0$, $\tilde{C_{11}}=\tilde{C_{22}}=C_{11}$. 
This non-isotropy behavior could be the outcome of the anharmonicity.

\section{Conclusion}
In summary, we introduced a method to obtain the continuum description of the elastic properties of monolayer h-BN
through {\em ab initio} density functional theory.
This thermodynamically rigorous continuum description 
of the elastic response is formulated by expanding the
elastic strain energy density in a Taylor series in strain 
truncated after the fifth-order term. 
we obtained a total of fourteen nonzero independent 
elastic constants for the up to tenth-order tensor. 
We predicted the pressure dependent second-order elastic moduli.
This continuum formulation is suitable for 
incorporation into the finite element method.

\begin{acknowledgments}
The authors would like to acknowledge the generous financial support from the Defense Threat Reduction Agency (DTRA) Grant \# BRBAA08-C-2-0130.
\end{acknowledgments}

\bibliography{hbn}

\end{document}